\documentclass[aps,preprintnumbers,amsmath,amssymb,superscriptaddress,letter]{revtex4}
\usepackage{graphicx}
\usepackage{subfigure}
\usepackage{color}
\usepackage{hyperref}
\newcommand{\bea}{\begin{eqnarray}}
\newcommand{\eea}{\end{eqnarray}}
\newcommand{\p}{\partial}

\usepackage{ulem} %

\begin{document}

\title{Violation of the quantum null-energy condition in a holographic wormhole and infrared effects}


\author{Akihiro {\sc Ishibashi}}\email[]{akihiro@phys.kindai.ac.jp}
\affiliation{%
{\it Department of Physics and Research Institute for Science and Technology, Kindai University, Higashi-Osaka 577-8502, JAPAN
}}

\author{Kengo {\sc Maeda}}\email[]{maeda302@sic.shibaura-it.ac.jp}
\affiliation{%
{\it Faculty of Engineering,
Shibaura Institute of Technology, Saitama 330-8570, JAPAN}}

\author{Eric {\sc Mefford}}\email[]{eric.mefford@polytechnique.edu}
\affiliation{%
{\it Department of Physics, University of California, Santa Barbara, California 93106, USA
}}
\affiliation{{\it Centre de Physique Th\'eorique, \'Ecole Polytechnique, CNRS, F-91128, Palaiseau, FRANCE}}

\begin{abstract}
We examine the quantum null energy condition~(QNEC) for a $2+1$-dimensional conformal field theory~(CFT) at 
strong coupling in the background of a wormhole spacetime by employing the AdS/CFT correspondence. First, we 
numerically construct a novel $3+1$-dimensional vacuum AdS black hole solution with nontrivial topology, which 
is dual to a wormhole geometry connecting two flat universes. Although the bulk null energy condition~(NEC) is not violated,  
the NEC for the holographic stress-energy tensor is violated near the wormhole throat. Next, we investigate the entanglement entropy for a half-space anchored to the boundary wormhole throat. We propose a natural prescription for regularizing the IR divergent part of the entanglement entropy and show that the QNEC is violated at the throat. This is the {\it first} counterexample to the QNEC, indicating that IR effects 
are crucial.  
\end{abstract}

\maketitle



\section{Introduction}

The null energy condition~(NEC) is key to understanding the basic properties of spacetime structure. 
It holds for most physically reasonable classical fields and 
plays a crucial role in various theorems concerning singularities~\cite{singularity_theorem} and 
black hole~(BH) mechanics~\cite{Hawking1971}. 
%
However, as a local condition, the NEC can be violated~\cite{EGJ1965} when quantum effects are considered. 
As an improved condition, an (achronal) averaged null energy condition (ANEC) 
that integrates the NEC along a null geodesic was proposed and used 
in improved versions of singularity theorems~\cite{GrahamOlum2007,FewsterGalloway2011} and
topological censorship~\cite{FSW1993}.  

The quantum null energy condition (QNEC)\cite{Bousso:2015mna, Bousso:2015wca, Koeller:2015qmn} is a new alternative condition to the NEC 
which is nonlocal, as it involves the von Neumann entropy [or entanglement entropy (EE)] $S$ of quantum fields in some subregion $A$ of the spacetime considered.  
More concretely, the QNEC gives a lower bound for the null-null component of the stress-energy tensor $T_{kk}$ as  
\begin{align}
\label{QNEC}
2\pi\int_{{\partial A}} \sqrt{\gamma}\,T_{kk}\ge \frac{D^2S}{D\lambda^2}, 
\end{align}  
where $D^2S/D\lambda^2$ is the second variation under null deformations of the von Neumann entropy for $A$, 
and $\gamma$ is the determinant of the boundary metric on the subregion boundary $\partial A$. 
The QNEC was originally derived from the quantum focussing conjecture~\cite{Bousso:2015mna}, 
where a ``quantum expansion" of the null geodesic congruence never increases toward the future. 
The QNEC was first shown in Minkowski spacetime for free bosonic field theories~\cite{Bousso:2015wca} and later for the cases of holographic CFTs in 
Minkowski space~\cite{Koeller:2015qmn} or in a class of curved spacetimes~\cite{Fu:2017evt}.

In this paper, we study the QNEC for quantum field theories at strong coupling on a wormhole geometry 
via the AdS/CFT correspondence~\cite{Maldacena:1997re} and show that it can be violated. Recently, bulk wormholes have gained some attention in the context of the AdS/CFT duality due to puzzles they raise when a bulk geometry connects multiple boundaries 
that each allow a well-defined QFT~\cite{MaldacenaMaoz04,GJW2017}. Here, we numerically construct a novel $3+1$-dimensional static-vacuum-AdS black hole solution with non-trivial 
topology, where the AdS {\it boundary} metric is conformal to a wormhole geometry that connects two flat universes. 
According to the AdS/CFT dictionary~\cite{Maldacena:1997re}, this corresponds to a thermal 
state in the boundary field theory at strong coupling on this background. As in the case of Ref.~\cite{GJW2017}, $T_{kk}$ is negative 
near the wormhole throat. We focus on the von Neumann entropy for a half-space subregion whose boundary is the wormhole 
throat. According to the HRT formula~\cite{Ryu:2006bv, Hubeny:2007xt}, the corresponding minimal surface is anchored to the AdS boundary at the throat and extends to the bulk black hole at spatial infinity. Because the half-space minimal surface asymptotically approaches the IR region of the black hole horizon, it shares the same IR divergence, and we propose a novel definition of the IR-regularized entropy $S_{reg}$ for the half-space:
\begin{align}
S_{reg} := S_{UV} - \frac{1}{2}S_{BH} \,. 
\label{S:reg}
\end{align} 
Here, $S_{UV}$ denotes the UV-regularized entropy for the given half-space and $S_{BH}$ denotes the Bekenstein-Hawking entropy of the bulk black hole. By this definition, we can subtract the thermal part of the entropy, and thereby manifest a purely entanglement part of the entropy for the boundary wormhole. We find that although Eq.~(\ref{QNEC}) is satisfied in a UV expansion near the AdS boundary, it can be violated when IR-effects near spatial infinity are taken into account. As far as we know, this is the first counterexample to the QNEC due to IR-effects.

We note that Fu, Koeller, and Marolf \cite{Fu:2017lps} considered an example that violates the QNEC in a curved spacetime within $d \geqslant 5$ Gauss-Bonnet theory under a local stationarity condition.  In \cite{Fu:2017evt}, the same authors proposed UV conditions for a curved spacetime QNEC to be preserved. In fact, under the new conditions, they showed that the QNEC in the same $d \geqslant 5$ Gauss-Bonnet theory is preserved. The UV conditions discussed in \cite{Fu:2017evt} are dependent upon the spacetime dimensions, and for $3$-dimensional curved spacetimes, the conditions simply reduce to local stationarity at the entangling surface. Our counterexample satisfies the local stationarity condition \cite{Fu:2017evt} at the wormhole throat, nevertheless the QNEC is violated.

In the next section, we numerically construct our bulk geometry with a three-dimensional wormhole on the conformal
boundary. We also give the regularized stress-energy tensor for CFT on the boundary wormhole and check that the
NEC itself is violated near the wormhole throat as expected. In section III, we holographically define our regularized entanglement entropy $S_{reg}$ for a half-space of the wormhole geometry by introducing the notion of a regularized
surface area $A_{reg}$. Then, in section IV, we numerically examine the QNEC in our wormhole spacetime and show that the QNEC is violated. Section V is devoted to a summary and discussion.

\section{The Bulk Geometry}
We first recall the $3+1$-dimensional static-vacuum-AdS black hole metrics for different horizon topologies, in units where $L_{AdS}=1$ and the conformal boundary is at $z=0$~\cite{Birmingham:1998nr, Emparan:1999gf}:
\begin{align}
ds^2 &= \frac{1}{z^2}\left[-f_k(z)dt^2 + \frac{dz^2}{f_k(z)} + d\Sigma_{k}^2\right],\nonumber\\
d\Sigma_{k}^2 &=\begin{cases} dr^2 + \sin(r)^2d\phi^2 \quad &k=1,\\ dr^2+r^2d\phi^2 \quad &k=0,\\ dr^2 + \cosh^2rd\phi^2 \quad &k=-1,\end{cases}\nonumber \\f_k(z) &=1+k\,z^2  - \mu\,z^3.
\label{AdSblackholes}
\end{align}
Here, $\mu$ determines the black hole mass and $k$ the horizon topology. For $k=1$, $r\in [0,\pi)$, for $k=0$, $r\in [0,\infty)$ and for $k=-1$, $r\in (-\infty,\infty)$.  Near the throat of the hyperboloid at $r=0$, the boundary metric is conformal to a cylinder. A static spacetime interpolating between a $k=-1$ black hole near $r=0$ and a $k=0$ black hole as $r\to\infty$ will have, on the boundary, a wormhole connecting two flat universes.

\begin{figure}[t!]
\begin{center}
\includegraphics[width=.4\textwidth]{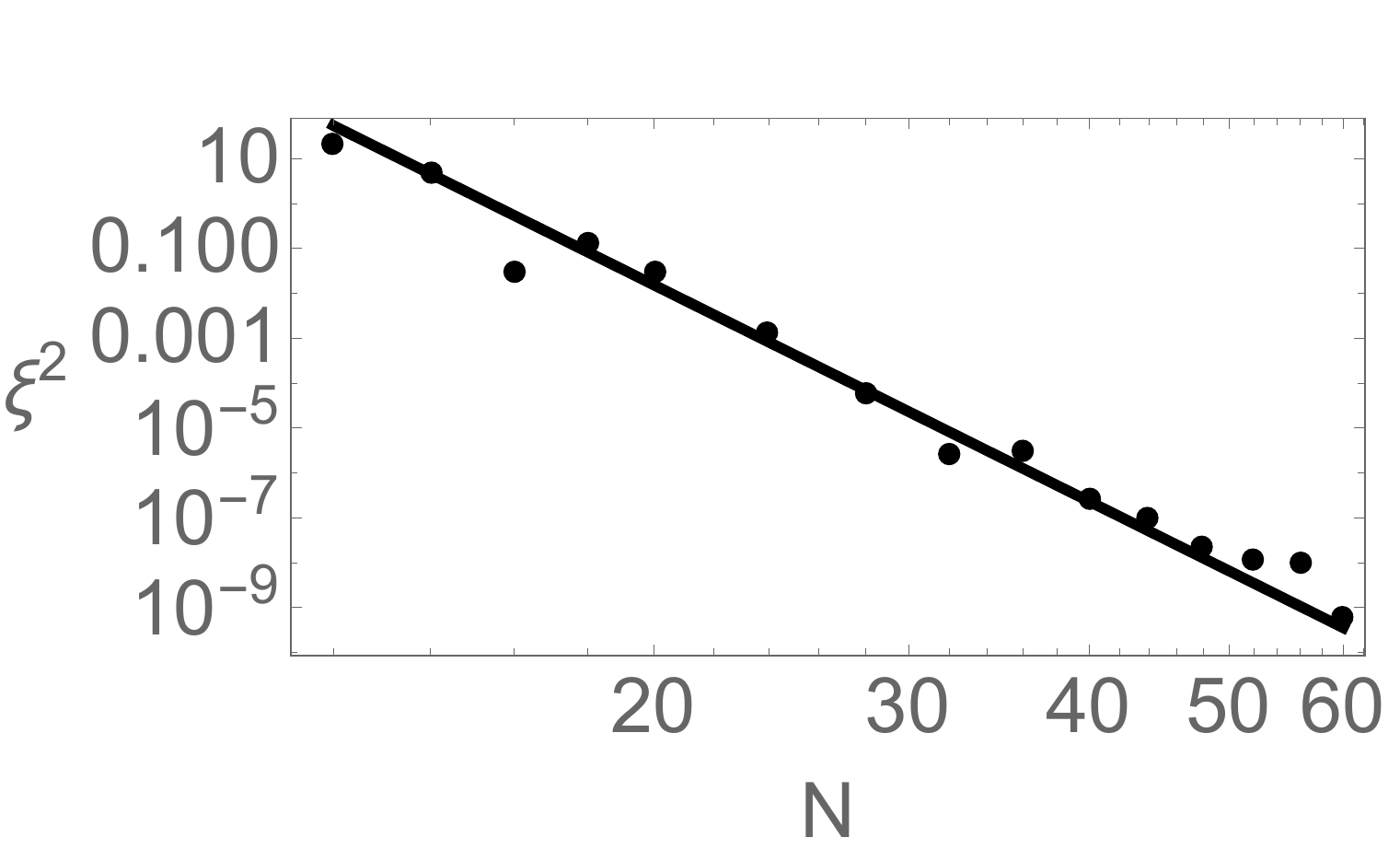}
\caption{\label{convergencefigure} Convergence of max($\xi^2$) for $N\times N$ grid and $\zeta=-.9$. Data presented below has $N=64$.}
\end{center}
\end{figure}

To find such a spacetime, we numerically solve the Einstein-DeTurck equation,
\begin{align}
R_{\mu\nu} + 3 g_{\mu\nu} - \nabla_{(\mu}\xi_{\nu)} = 0.
\label{EinsteinDeTurck}
\end{align}
The DeTurck vector is defined in terms of the Christoffel symbols of the metric $g$ and a reference metric $\bar{g}$ as $\xi^{\mu} = g^{\alpha\beta}\left(\Gamma^{\mu}_{\alpha\beta} - \bar{\Gamma}^\mu_{\alpha\beta}\right)$. The final term in Eq.~(\ref{EinsteinDeTurck}) fixes a gauge for the vacuum Einstein equation resulting in a set of elliptic rather than hyperbolic equations better suited to numerical boundary value problems \cite{Headrick:2009pv,Figueras:2011va}. Solutions to Eq.~(\ref{EinsteinDeTurck}) with $\xi^\mu = 0$ also solve the Einstein equation. Subject to certain boundary conditions, Eq.~(\ref{EinsteinDeTurck}) satisfies a maximum principle; hence, an appropriate choice of $\bar{g}$ will ensure that $\xi^\mu$ vanishes everywhere \cite{Figueras:2011va}. 
We use $\xi^2$ as a test of numerical accuracy, as shown in Fig.~\ref{convergencefigure}.

We choose a numerical domain of  $x\in [0,1]$ and $y\in [0,1]$ in which we construct one half of the wormhole geometry. The other half follows from a reflection across $x=0$. Over this domain, an ansatz for $g$ (modified from Ref.~\cite{Santos:2012he}) is
\begin{align}
ds^2 &= \frac{1}{g(x)^2y^2}\biggl[-(1-y)f(x,y)Tdt^2 + \frac{g(x)^2A}{(1-y)f(x,y)}dy^2
+ \frac{4B(dx+x(1-x^2)^2Fdy)^2}{(1-x^2)^4}+\frac{\ell(x)S}{(1-x^2)^2}d\phi^2\biggr]
\end{align}
where $0\le \phi\le 2\pi$, $X\equiv\{T, S, A, B, F\}$ are functions of $x$ and $y$ and
\begin{align}
f(x,y) &= 1+y+y^2x^2(3-2x^2),\nonumber
\\ g(x)&= 2+x^2(3-2x^2),
\\ \ell(x) &= \zeta(1-x^2)^4+[1+x^2(1-x^2)^2]^2.\nonumber
\end{align}
We choose $T=S=A=B=1$ and $F=0$ for $\bar{g}$. Using a Newton-Raphson pseudospectral numerical method over an $N\times N$ Chebyshev grid also requires a seed for which we chose $\bar{g}$. We impose the boundary conditions
\begin{align}
\partial_x T|_{x=0}&=\partial_x S|_{x=0}=\partial_x A|_{x=0}=\partial_x B|_{x=0}=\partial_x F|_{x=0}= 0, \nonumber\\
T|_{x=1}&=S|_{x=1}=A|_{x=1}=B|_{x=1}=1,\quad F|_{x=1}=0,\\
T|_{y=0}&=S|_{y=0}=A|_{y=0}=B|_{y=0}=1,\quad F|_{y=0}=0,\nonumber\\
&\quad\quad\quad\quad\quad T_{y=1}=A|_{y=1}.\nonumber
\label{boundaryconditions}
\end{align}

A smooth, constant temperature horizon requires $\frac{T}{A}|_{y=1}=1$, giving the last boundary conditions. Further boundary conditions at $y=1$ are easily found from the near horizon expansions $X(x,y) \approx X(x,1) + (y-1)\partial_y X(x,y)|_{y=1} + ...$ into Eq.~(\ref{EinsteinDeTurck})~\cite{ILL}. For all $\zeta$, the bulk black hole has temperature $T_{BH} = \frac{1}{4\pi}$.

\begin{figure}[t!]
\begin{center}
\includegraphics[width=.4\textwidth]{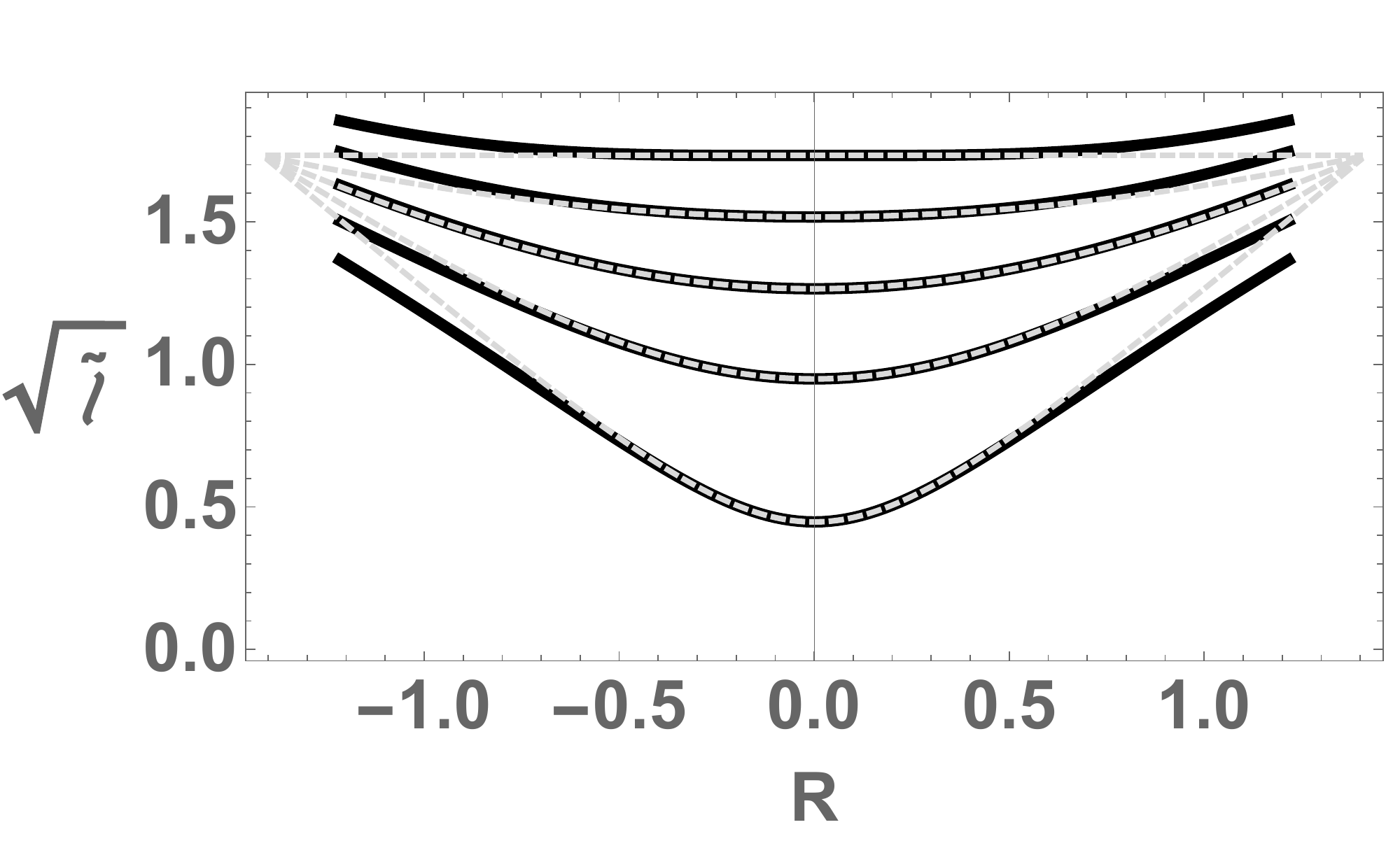}
\includegraphics[width=.4\textwidth]{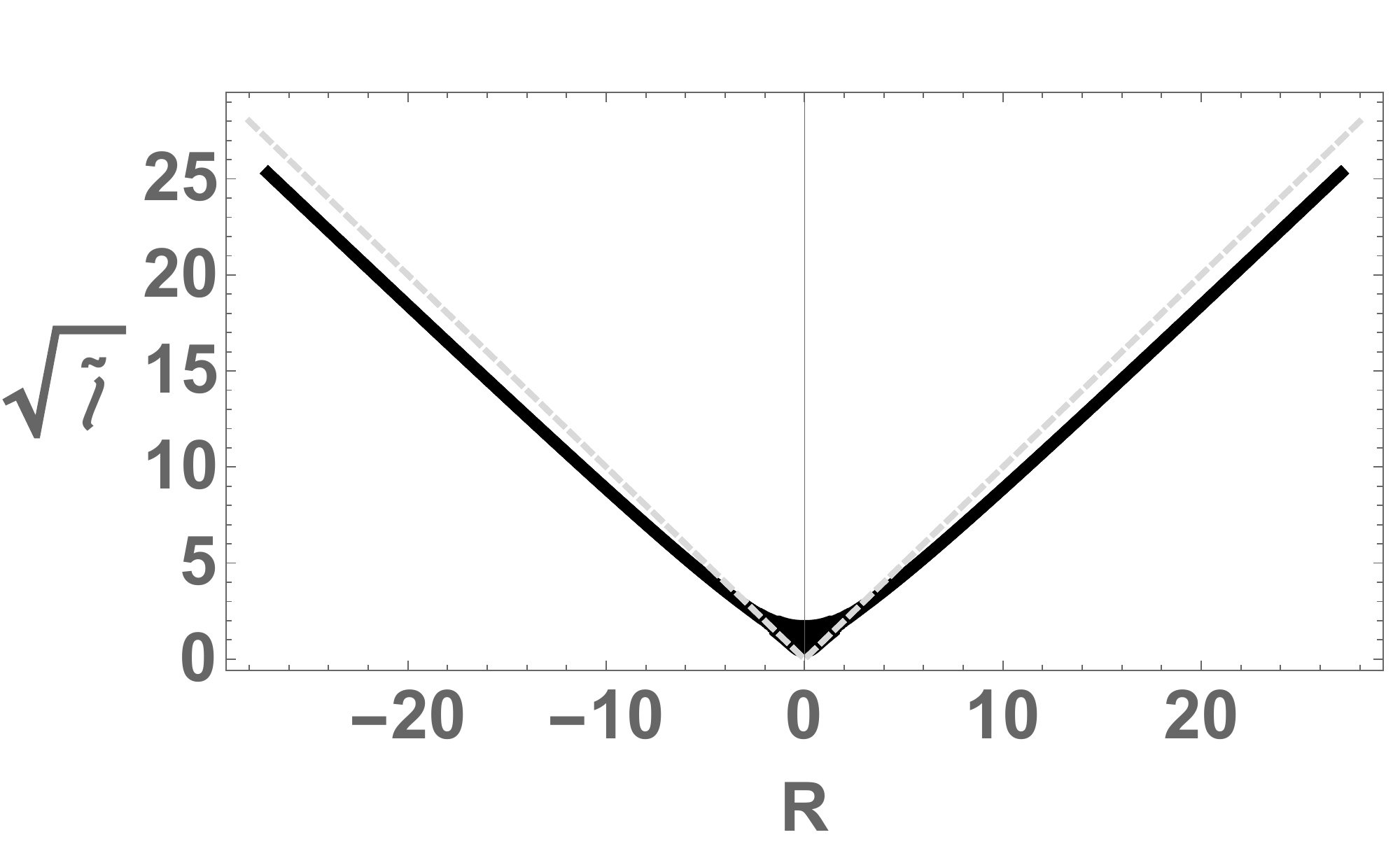}
\caption{\label{wormholeradiusfigure} The wormhole radius as a function of the boundary coordinate $R$. (Left) The radius near the wormhole throat (black) against $\sqrt{1+\zeta +(1-\frac{\zeta}{2})R^2}$ (gray dotted) for $\zeta = -.8$ to $\zeta=2$ in intervals of $.7$. (Right) The large $\sqrt{\tilde{\ell}(R)}$ radius plotted against $R$ for the same values of $\zeta$.}
\end{center}
\end{figure}

If, at $x=0$, we instead chose to impose Dirichlet conditions, $T=S=A=B=1,\, F=0,$ as well as the redefinition $t\mapsto2t$, we would explicitly impose the $k=-1$ metric in Eq.~(\ref{AdSblackholes}). However, all that is required for our wormhole geometry is the reflection symmetry. Furthermore, at $x=1$, the redefinitions $x= \sqrt{1-\frac{1}{3R}}, t\mapsto3t$ give the $k=0$ metric in Eq.~(\ref{AdSblackholes}). At $y=0$, with $R \equiv \int^x \frac{2dx'}{(1-(x')^2)^2}$,
 the metric is
\begin{align}
ds^2=\frac{dy^2}{y^2}+\frac{1}{y^2g^2(x)}\left[-dt^2 + dR^2 + \tilde{\ell}(R)d\phi^2\right].
\end{align}
Near $x=0$, $\tilde{\ell}(R) = 1 + \zeta + (1-\frac{\zeta}{2})R^2 +\mathcal{O}(R^4)$ and near $x=1$, $\tilde{\ell}(R) = R^2 + \mathcal{O}(R)$. 
When $-1<\zeta<2$, the $S^1$ is minimized at $R=0~(x=0)$ corresponding to the throat of a wormhole 
connecting two flat universes. This surface is locally stationary for any $\zeta$. Plots of the wormhole radius as a function of $R$ and $\zeta$ are shown in Fig.~(\ref{wormholeradiusfigure}).

The spacetime is asymptotically locally AdS \cite{Fischetti:2012rd} with Fefferman-Graham~(FG) expansion near $z=0$,
\begin{align}
ds^2 = \frac{1}{z^2}\left[dz^2 + (h^{(0)}_{ab} + z^2h_{ab}^{(2)} +z^3 h_{ab}^{(3)}+...)dx^a dx^b\right].
\label{FGexpansion}
\end{align}
The following expansions, inserted into Eq.~(\ref{EinsteinDeTurck}) and solved order by order in $y$, can be used to find $h^{(i)}_{ab}$:
\begin{align}
X(x,y) &\approx  X^{(0)}(x) + X^{(2)}(x)\frac{y^2}{2} + X^{(3)}(x)\frac{y^3}{6} + \mathcal{O}(y^4)\nonumber\\
y(z,r) &\approx z\left[ \frac{1}{g(r)} + z^2 y^{(3)}(r) + z^3 y^{(4)}(r) + \mathcal{O}(z^4)\right]\nonumber\\
x(z,r) &\approx r+ z^2 x^{(2)}(r) + z^4 x^{(4)}(r)+\mathcal{O}(z^5).
\label{FGexpansion2}
\end{align}
Analytic expressions exist for $X^{(2)}$ in terms of geometric invariants on the boundary, 
but $X^{(3)}$ require numerics \cite{deHaro:2000vlm}. The expressions are not illuminating and are omitted. 
An example of this procedure is in Ref.~\cite{Mefford:2016res}. From this expansion, one finds that the regularized holographic stress-energy tensor is \cite{deHaro:2000vlm}:
\begin{align}
\langle T_{ab} \rangle = \frac{3h_{ab}^{(3)}}{16\pi G_4}.
\label{regularizedstresstensor}
\end{align}
The off-diagonal terms vanish and the diagonal terms are
\begin{align}
h_{ij}^{(3)}dx^idx^j =\frac{1}{6g(r)^3}\left[h_{tt}^{(0)}(T^{(3)}-2j) dt^2 + h_{rr}^{(0)}(B^{(3)}+j) dr^2 + h^{(0)}_{\phi\phi}(S^{(3)}+j) d\phi^2 \right],\quad j=2g(r)-4.
\end{align}
Importantly, we find $T^{(3)}+S^{(3)}+B^{(3)}=0$ implying a vanishing trace of the stress-energy tensor, as required for a CFT$_3$. 

\begin{figure*}[t!]
\includegraphics[width=.36\textwidth]{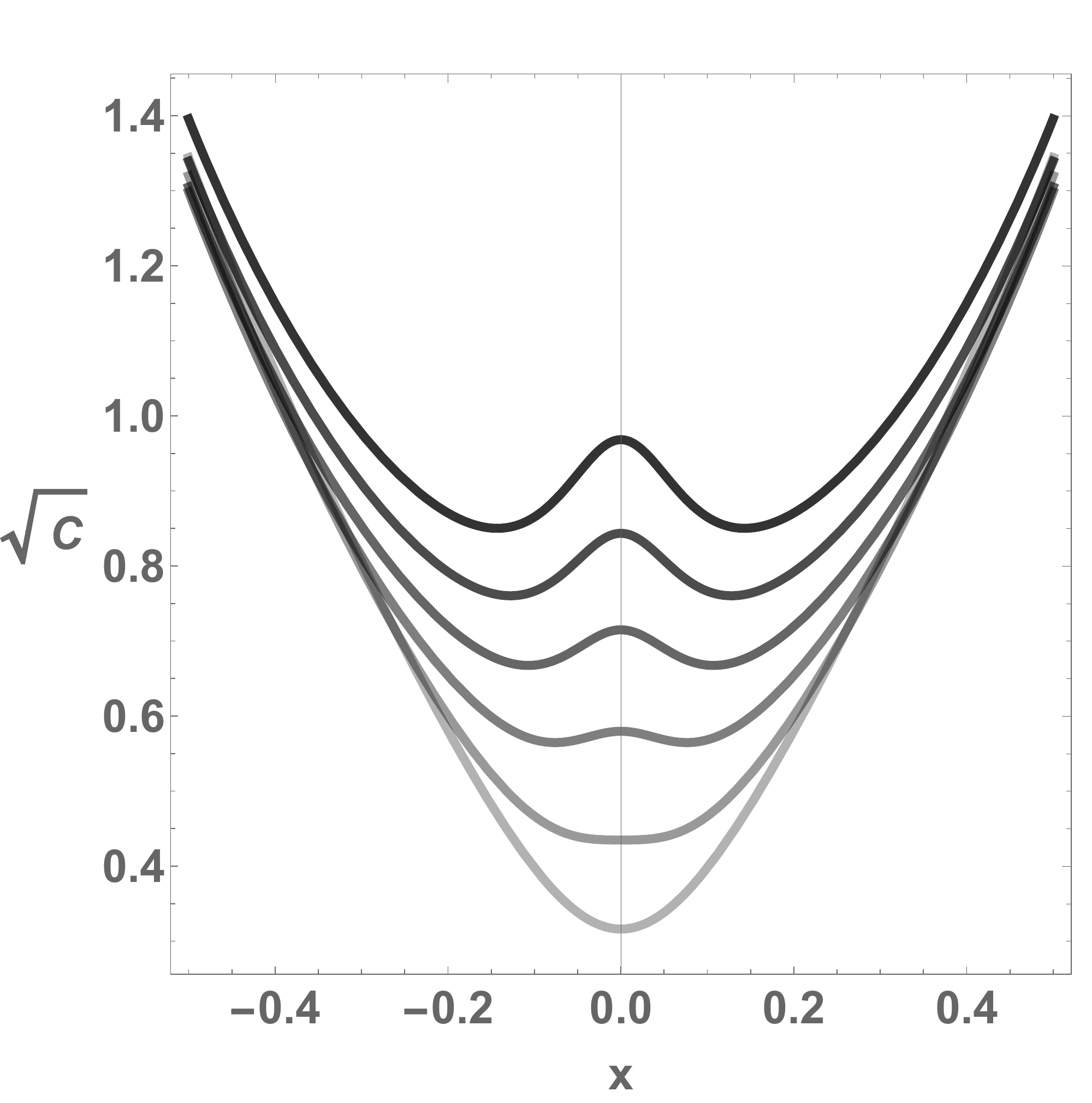}
\includegraphics[width=.4\textwidth]{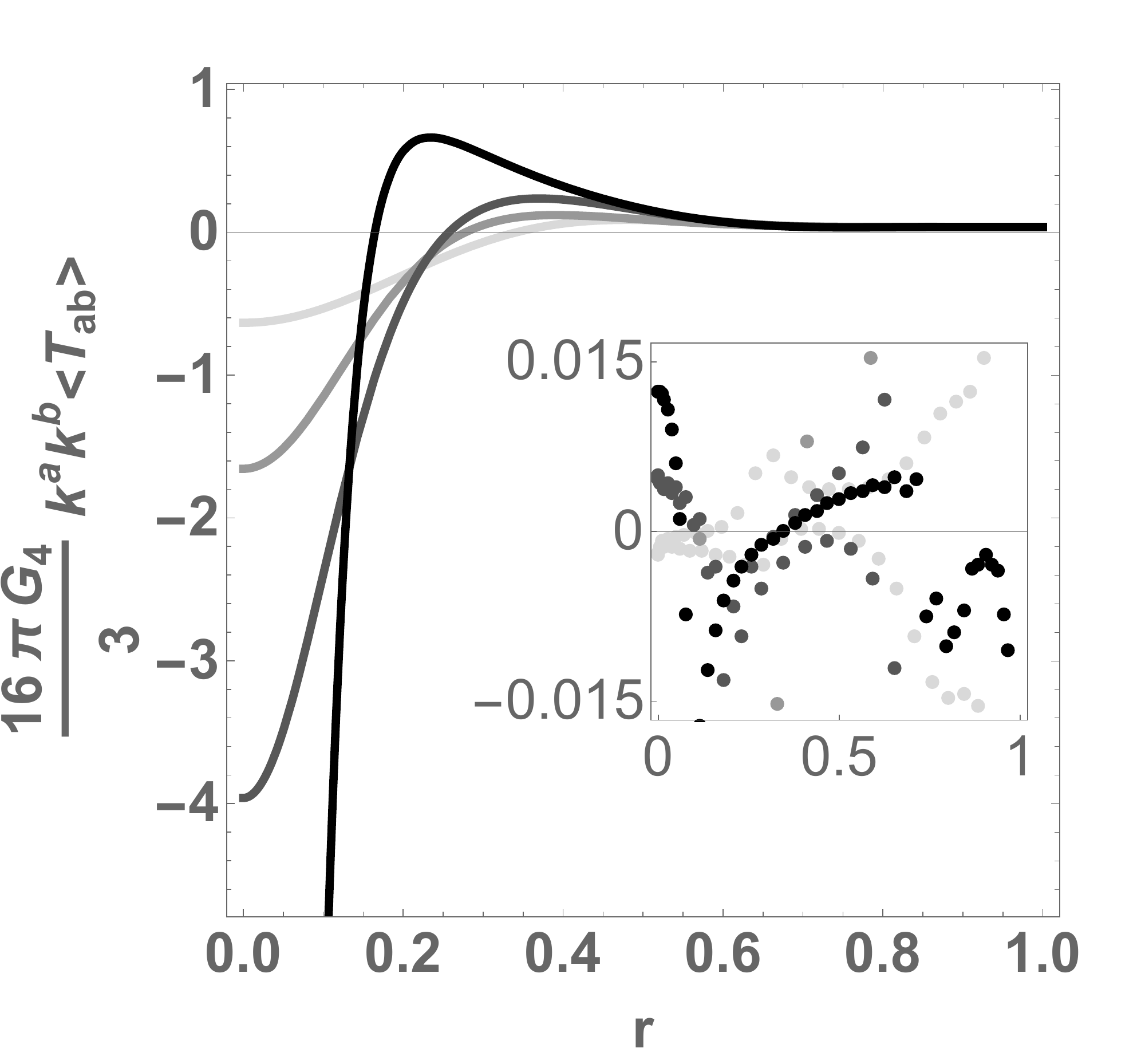}
\caption{\label{nullstressenergy}(Left) The radius of the wormhole as a function of bulk position for $\zeta=-.9$. Here, $C=\ell(x)S(x,y)/(1-x^2)^2$. Curves are shown for $y=0,.2,.4,.6,.8,1$ (light to dark). (Right) Null energies for $\zeta = -.9, \,-.6,\,-.4, \,0$ (dark to light). (Inset) Estimating the numerical error, $\frac{B^{(3)}+T^{(3)}+S^{(3)}}{\sqrt{(B^{(3)})^2+(T^{(3)})^2+(S^{(3)})^2}}$ (same shading).}
\end{figure*}

The null-null component of the stress-energy tensor, plotted in Fig.~\ref{nullstressenergy}, can be defined in terms of the future directed null vector $k^a = \p_t + \p_{R}$ with $R$ defined as before: $R \equiv \int dr \sqrt{h_{rr}^{(0)}}$. For $-1<\zeta<2$, 
$\langle T_{kk}\rangle < 0$ near the throat. Near spatial infinity, $\langle T_{kk} \rangle =\frac{3}{16\pi G_4}(\frac{1}{27})>0$, which is the value for a $T_{BH}=\frac{1}{4\pi}$, $k=0$ black hole divided by 27 due to the choice of conformal frame [see Eq.~(\ref{FGexpansion2})]. 

\begin{figure*}[t!]
\subfigure[\label{minimalsurfaces}]{\includegraphics[width=.305\textwidth]{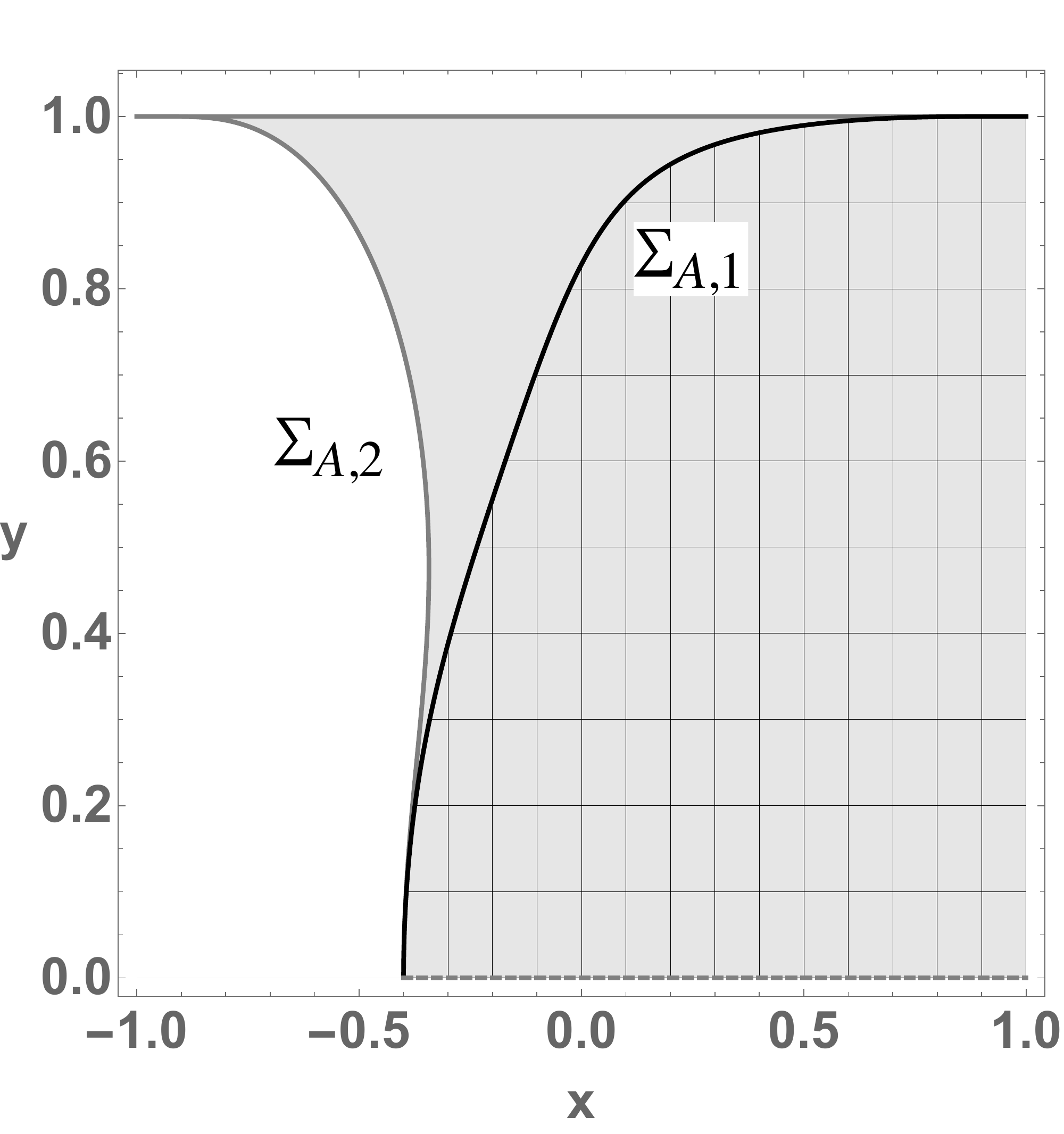}}
\subfigure[\label{regularizedareas}]{\includegraphics[width=.33\textwidth]{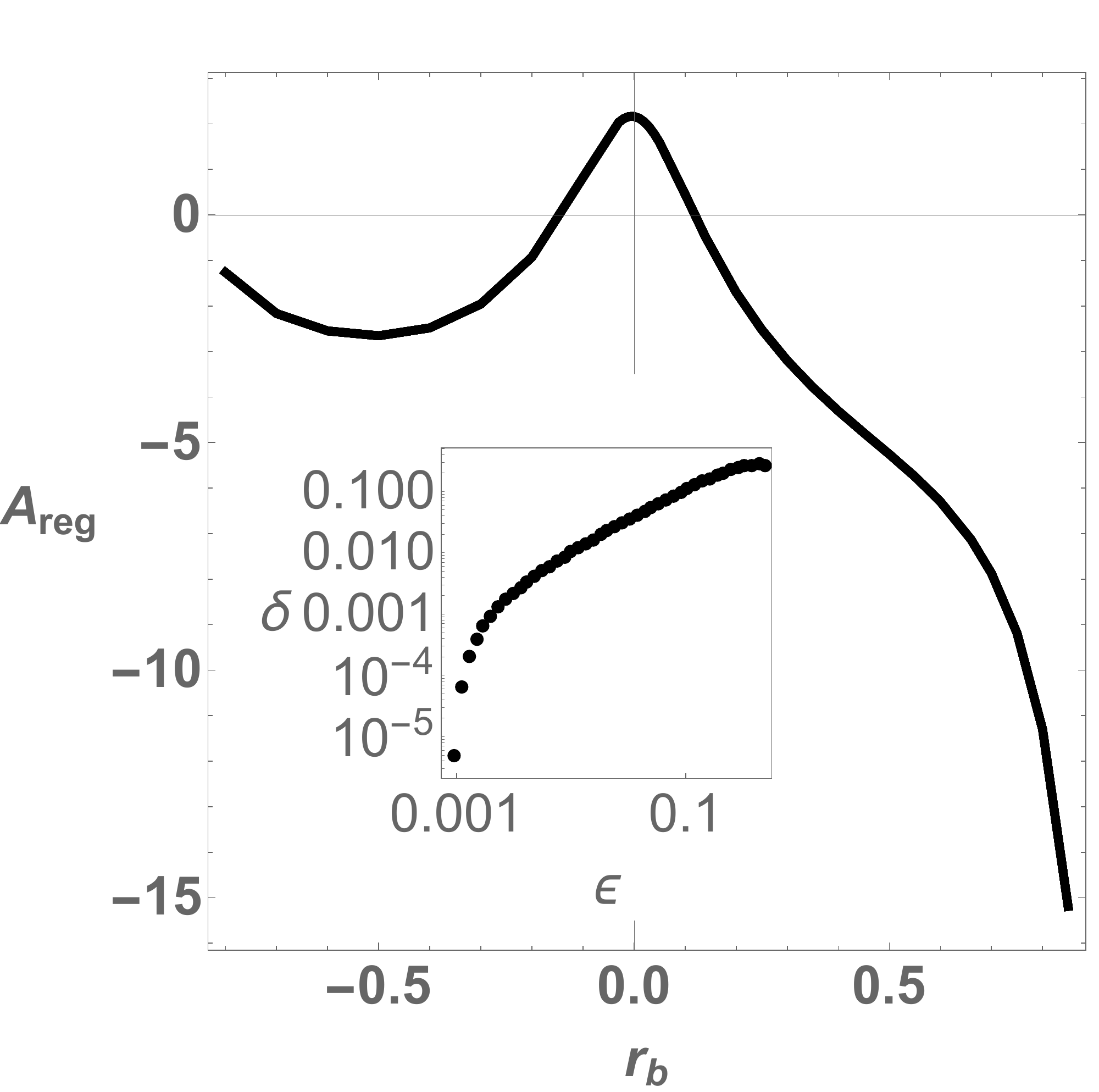}}
\caption{(a) $\Sigma_{A,1}$ (black) and $\Sigma_{A,2}$ (gray) for $r_{b}=-.4$ and $\zeta=-.9$. Filled/hatched regions illustrate the homology constraint. (b) $\mathcal{A}_{reg}$ for $\zeta=-.9$. (Inset) Estimating numerical error, $\delta = |\frac{\mathcal{A}_{reg}(\epsilon)}{\mathcal{A}_{reg}(\epsilon_*)}-1|$ for $\epsilon_* = 10^{-3}$ and $r_{b}=.5$.  }
\end{figure*}

\section{Half-space Entanglement Entropies} We are interested in the entanglement entropy of the subregion $A$ defined by $r_b\le r< \infty$. We call this a ``half-space'' entanglement entropy since $\p A$ at $r=r_b$ splits the fixed time $t$ Cauchy surface 
into two pieces (if $r_b=0$, this is the wormhole throat). The holographic entanglement entropy of the subregion $A$ is given by the HRT formula \cite{Ryu:2006bv, Hubeny:2007xt},
\begin{align}
S(A) = \frac{\mathcal{A} (\Sigma_A)}{4G_4},
\end{align}
where $\Sigma_A$ is the codimension-2 minimal area surface in the bulk anchored to the boundary at $\partial A$ and homologous to $A$ with 
${\cal A}(\Sigma_A)$ denoting its area~\cite{Headrick:2007km}. As shown in Fig.~\ref{minimalsurfaces}, there are two competing minimal surfaces. One, which we call $\Sigma_{A,1}$, touches the black hole horizon as $x\to 1$, and the other, which we call $\Sigma_{A,2}$, touches the black hole horizon as $x\to -1$ (the reflection of $\Sigma_{A,1}$ for $r\geq -r_{b}$) and also includes the black hole horizon, as required by the homology constraint. Notably, the areas of both surfaces diverge due to the noncompact black hole horizon with finite cross-sectional area. However, since $\Sigma_{A,1}$ diverges in only one direction, its area is always less than $\Sigma_{A,2}$. In particular, this means that the minimal surfaces never undergo a phase transition. From here on, we will refer to $\Sigma_{A,1}$ as $\Sigma_{A}$, the minimal surface for the region $A$.

To find $\Sigma_{A}$, we minimize the area functional
\begin{align}
{\cal A} = 2\pi \int_0^1 ds \sqrt{g_{\mu\nu}\partial_s Y^\mu \partial_s Y^\nu}
\label{areaintegral}
\end{align}
where $Y^\mu(s) = \{x(s),y(s)\}^\mu$. The minimal surfaces have boundary conditions $\{x,y\}|_{s = 0} = \{r_{b},0\}$ and $\{x, \partial_{s}y\}|_{s=1}=\{1,0\}$ and solutions have $y|_{s=1}=1$.

\begin{figure*}[t!]
\includegraphics[width=.38\textwidth]{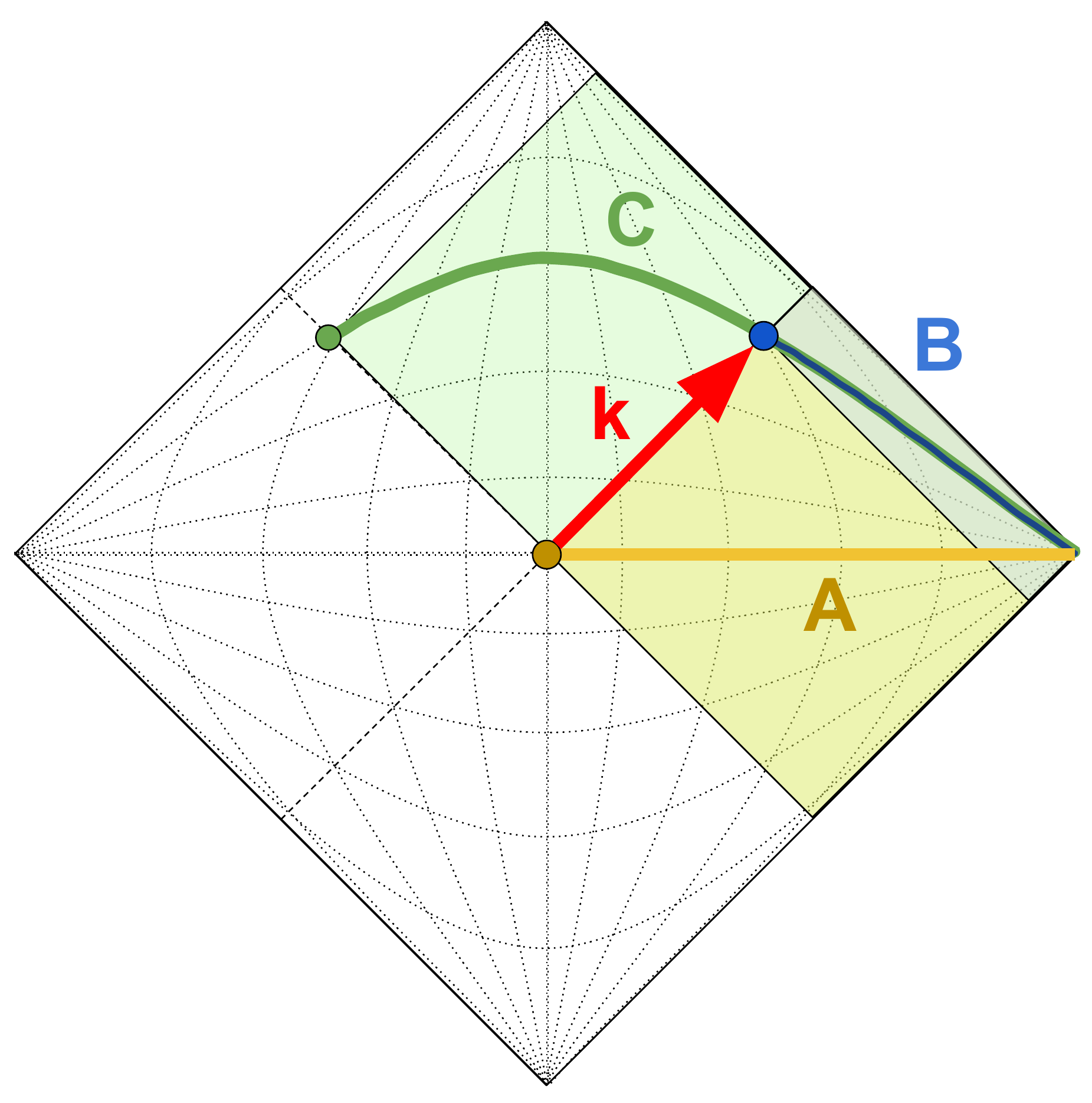}
\caption{\label{Penrose}In static AF universes, half-space subregions and their minimal surfaces lie on a single $t$ slice (horizontal lines). $\mathcal{D}(B)\subseteq \mathcal{D}(A) \subseteq \mathcal{D}(C)$ (shaded) implies $\Sigma_A, \Sigma_B, \Sigma_C$ are achronally separated.}
\end{figure*}

Eq.~(\ref{areaintegral}) is both UV and IR divergent. From the boundary perspective, the divergences in the former case are due to short distance correlations across $\partial A$ and in the latter case to thermal correlations extending to spatial infinity. To eliminate the UV divergence, we introduce a bulk UV cutoff for the integral Eq.~(\ref{areaintegral}), $z(x,y)= \epsilon$. 
As usual, the UV-part can be regularized by subtracting a counterterm proportional to the area of $\partial A$ as
\begin{align}
{\cal A}_{UV}:={\cal A} - \frac{2\pi \sqrt{\ell(r_{b})}}{g(r_{b})(1-r_{b}^2)}\frac{1}{\epsilon}.
\end{align}
In the bulk, the IR divergence comes from the $\{x,y\}\to\{1,1\}$ region where the minimal surface and bulk black hole coincide. Hence, the area of the minimal surface has the same divergence as half the black hole. As a concrete realization of Eq.~(\ref{S:reg}), we define a regularized area, 
${\cal A}_{reg}$ as
\begin{align}
{\cal A}_{reg} := {\cal A}_{UV} - \frac{1}{2}{\cal A}_{BH} \,. 
\label{regularizedentropy}
\end{align} 
A plot of $\mathcal{A}_{reg}$ is shown in Fig.~\ref{regularizedareas}.

As a simple example that will be useful in understanding the QNEC, we note the half-space minimal surface for a static cylindrical black hole where $d\Sigma_0^2 = dx_1^2 + dx_2^2$ in Eq.~(\ref{AdSblackholes}) with $x_2 = x_2+L_2$. With fixed $x_1$ subregion boundaries, the minimal surfaces solve the equation
\begin{align}
x_1'(z) = \frac{(c_1z)^2}{\sqrt{(1-(c_1z)^4)(1-\mu z^3)}}.
\label{compactifiedblackbrane}
\end{align}
For $|c_1|>\mu^{1/3}$, the minimal surface has a turning point and gives the entanglement entropy for a strip-shaped subregion. Half-space subregions have $|c_1|\leq\mu^{1/3}$ and the surface with minimal area that touches the horizon at $x_1\to\infty$ has $|c_1|=\mu^{1/3}$. Furthermore, near $z=0$, the minimal surface satisfies a UV expansion
\begin{align}
x_1(z) = x_{\min} + \frac{c_1}{3}z^3 + \mathcal{O}(z^4).
\end{align}

\begin{figure*}
\subfigure[\label{QNECviolation}]{\includegraphics[width=.4\textwidth]{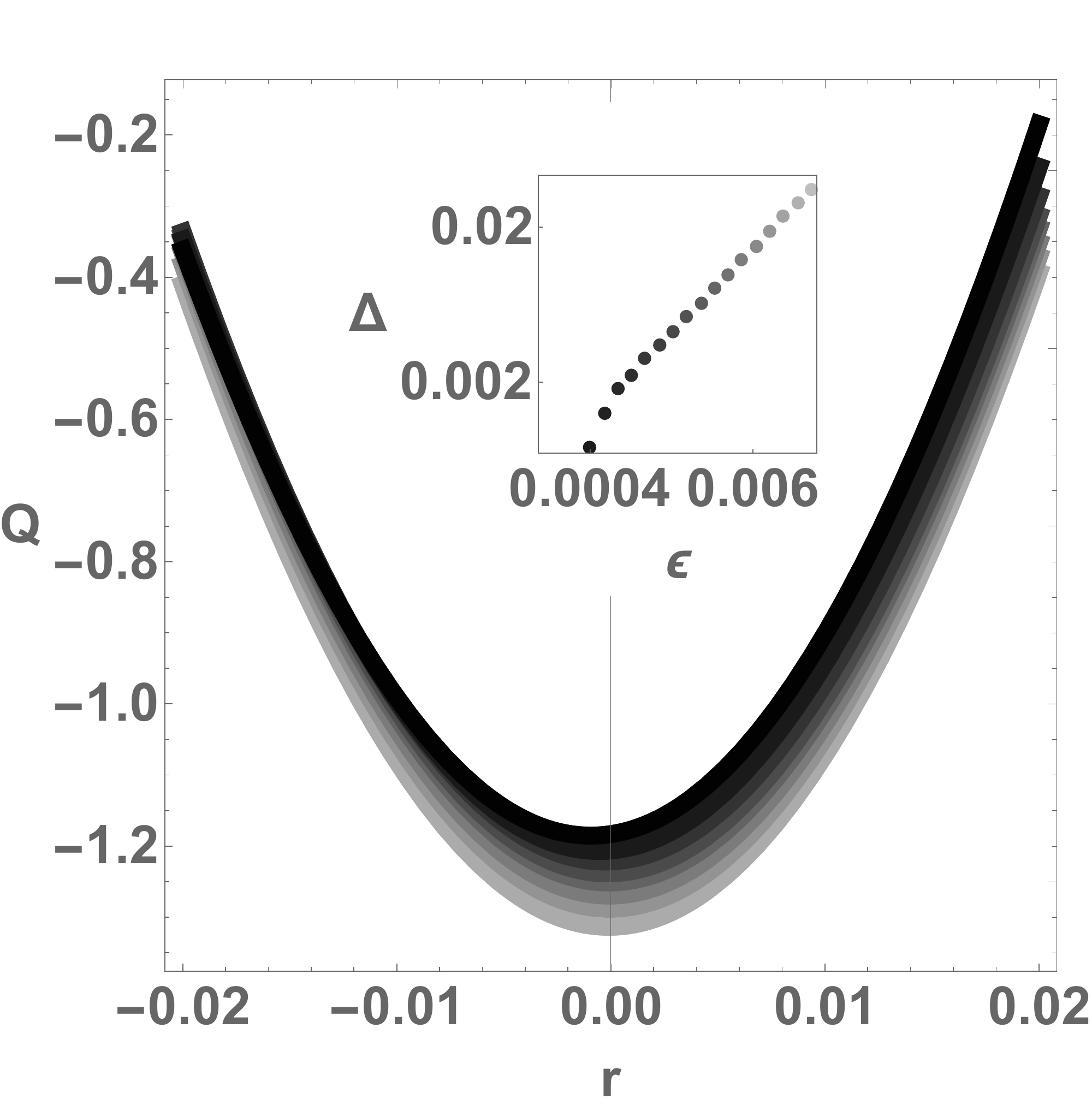}}
\subfigure[\label{UVdiscrepancy}]{\includegraphics[width=.4\textwidth]{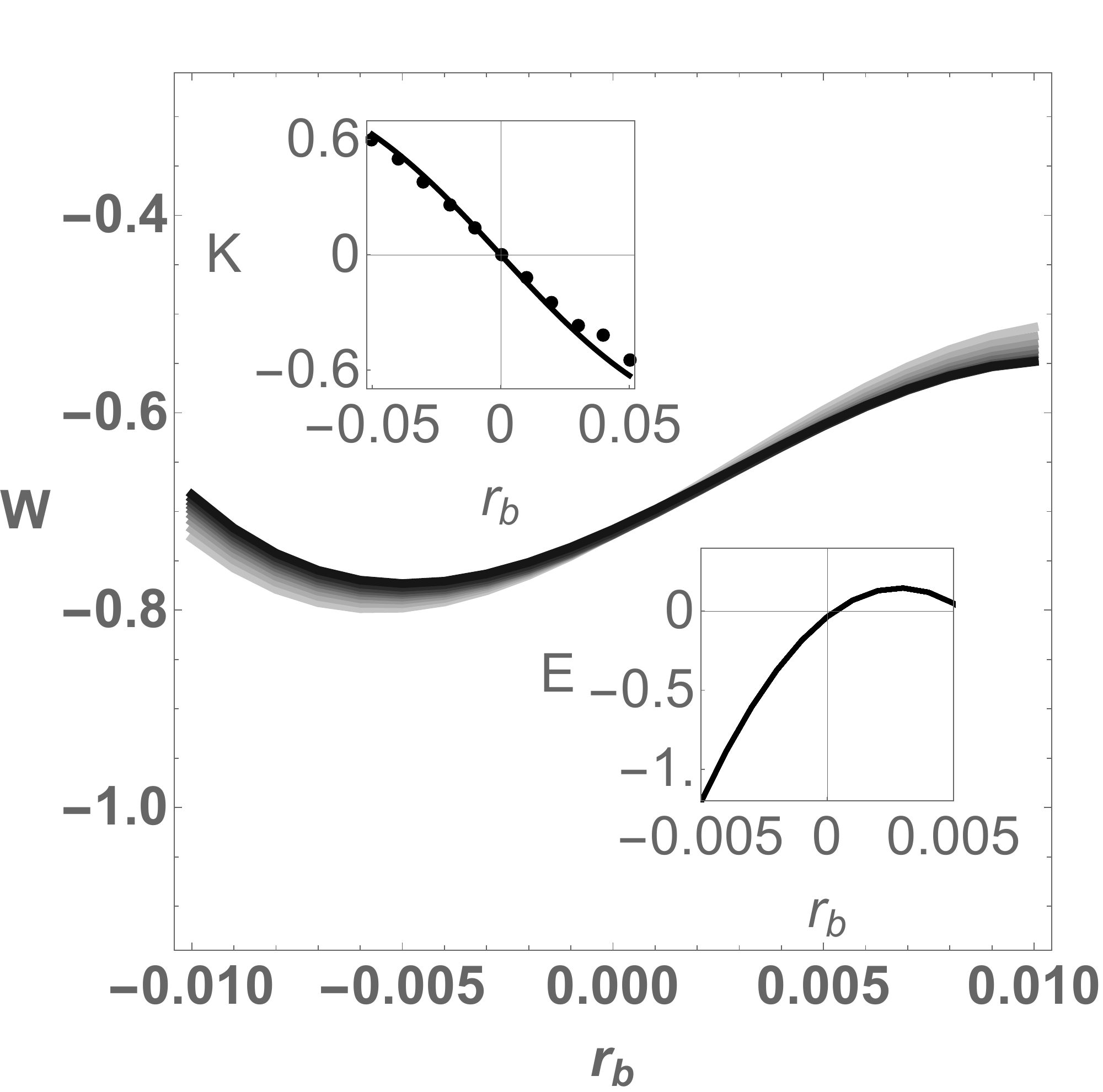}}
\caption{(a) Eq.~(\ref{QNEC2}) for $\zeta=-.9$. Curves correspond to $\epsilon$ from $.01$ to $10^{-4}$ (light to dark). (Inset) Estimating numerical error, $\Delta \equiv |\frac{Q(\epsilon)}{Q(\epsilon_*)}-1|$ with $\epsilon_*=5\times 10^{-4}$. (b) Non-vanishing of Eq.~(\ref{UVvariation}) due to IR effects. Curves correspond to $\epsilon$ from $.01$ to $10^{-4}$ (light to dark). (Inset, left) Comparison of analytic $K$ to numerical fits of the minimal surface embedding $\mathcal{O}(z^2)$ terms. (Inset, right) EWN is satisfied and nearly saturated.}
\end{figure*}

\section{Violation of the QNEC.}
The QNEC relates the null energy density of a QFT to the second null variation of the entanglement entropy at a point $p$ on $\partial A$. Though the individual pieces of the QNEC are UV divergent, given certain conditions at $p$, the combination in Eq.~(\ref{QNEC}) is finite. The condition for a $2+1$-dimensional curved spacetime is that the expansion $\theta|_{p}$ vanishes \cite{Fu:2017evt}. In our spacetime, this criteria is satisfied at the wormhole throat, $r_b=0$, independent of $\zeta$. 
An important point is that, {\it a priori}, the IR divergences of our minimal surfaces do not contribute to the QNEC as they are the same for all half-space entanglement entropies [see Eq.~(\ref{regularizedentropy})]. Said differently, the QNEC is manifestly IR finite since it involves a local variation of the entangling surface. 

 To simplify the numerics and exploit the isometries along $\partial_t$ and $\partial_\phi$, we investigate an integrated form of the QNEC where all points of $\partial A$ are moved equally in the $k^a$ direction, as in Eq.~(\ref{QNEC}). Importantly, because our spacetimes are static and asymptotically flat (AF), the half-space minimal surface lies on a single $t$ slice.  Hence, null and spatial variations of the entanglement entropy are proportional. This is shown in Fig.~\ref{Penrose}. For example, in the case of the static cylindrical black hole, the null variation of the half-space entanglement entropy vanishes, due to translation invariance in the $x_1$-direction. Importantly, this is not the case for strip-shaped subregions whose minimal surface will be time-dependent when the two subregion endpoints lie on different $t$ slices.

In terms of regularized quantities and our wormhole metric, Eq.~(\ref{QNEC}) becomes \cite{derivation},
\begin{align}
Q\equiv 2\pi \langle T_{kk} \rangle - \frac{1}{32\pi G_4 \sqrt{1+\zeta}}\frac{\delta^2 \mathcal{A}_{reg}}{\delta r^2}\biggr|_{r=0}\geq0.
\label{QNEC2}
\end{align}
In Fig.~\ref{QNECviolation}, this inequality is violated at the throat.

The QNEC violation is surprising because many proofs exist \cite{Bousso:2015wca, Koeller:2015qmn,Fu:2017evt, Balakrishnan:2017bjg}. However, these proofs do not consider thermal states, where IR degrees of freedom play an important role. Proofs in Refs. \cite{Koeller:2015qmn,Fu:2017evt} rely on entanglement wedge nesting (EWN), a statement that for subregions $A$ and $B$, if the domains of dependence satisfy $\mathcal{D}(B)\subseteq\mathcal{D}(A)$, then $\Sigma_A$ and $\Sigma_B$ are achronally separated \cite{Wall:2012uf}. In Fefferman-Graham gauge, one can show that the minimal surface embedding satisfies a UV expansion, 
\begin{align}
x(z) = r_{b} + \frac{K}{2}z^2 + \frac{c_1}{3}z^3 + \mathcal{O}(z^4),
\end{align}
where $K$ is the trace of the extrinsic curvature at $r_{b}$ \cite{extrinsicdefinition}. This is true in our work. Entanglement wedge nesting constrains $\delta c_1/\delta r$ to be bounded by a function proportional to $\langle T_{kk} \rangle$. The main assumption of Refs. \cite{Koeller:2015qmn,Fu:2017evt} is that when $K = 0$ at a point $p$, we also have the vanishing of
\begin{align}
W\equiv\frac{1}{\sqrt{\gamma}}\frac{\delta {\cal A}(\Sigma_A)}{\delta r} + h^{(0)}_{rr}c_1\biggr|_{p}=0 \,.
\label{UVvariation}
\end{align}
Hence, we may replace $c_1$ with $\delta \mathcal{A}(\Sigma_A)/\delta r$ and another variation results in the QNEC. However, thermal states do not generically obey this equation. For a strongly-coupled thermal CFT on a cylinder (dual to the cylindrical black hole), ${\delta {\cal A}(\Sigma_A)} / {\delta r}=0$. This follows from the translation symmetry of the spacetime. However, $c_1$ vanishes only when $T_{BH}=0$ [$\mu \to 0$ in Eq.~(\ref{compactifiedblackbrane})]. Likewise, in Fig~\ref{UVdiscrepancy}, we show that in the wormhole, Eq.~(\ref{UVvariation}) does not hold. However, as shown in Fig.~\ref{UVdiscrepancy}, the constraint of entanglement wedge nesting on $c_1$~\cite{EWNexpression}, which is obeyed and agrees with Eq.~(\ref{QNEC2}) in the vacuum, is
 \begin{align}
 E \equiv 2\pi G_4 \langle T_{kk} \rangle + \frac{h^{(0)}_{rr}}{4}k^r \delta_\lambda c_1|_{r=0}\geq0.
 \label{UVQNEC}
 \end{align}
This inequality is not only obeyed but nearly saturated as has recently been conjectured~\cite{Leichenauer:2018obf}. That the constraint on on the UV parameter $c_1$ from entanglement wedge nesting is obeyed, but the QNEC is violated demonstrates that the entanglement variation in Eq.~(\ref{QNEC2}) crucially includes an IR contribution. Furthermore, whereas for the cylindrical black hole, the everywhere positive energy density realizes the QNEC, in the wormhole geometry, a negative energy at the throat leads to its violation.
 
\section{Summary and discussion}

The QNEC illustrates new and beautiful connections between interacting QFTs and gravity. Intriguingly, it relates the variation of a nonlocal observable, the entanglement entropy of a subregion, to a local observable, $ T_{kk} $, and is believed to hold even in curved spacetime at points where $\theta|_p$ vanishes [where Eq.~(\ref{QNEC}) and Eq.~(\ref{QNEC2}) are equal]. However, for thermal states, which have finite energy density that extends to spatial infinity, less is known. In this paper, we have demonstrated that for these states on a particular wormhole background, the QNEC is violated. On the other hand, 
a purely UV expression, Eq.~(\ref{UVQNEC}), is obeyed and nearly saturated. This is a hint that the QNEC in Eq.~(\ref{QNEC}) may govern states perturbatively close to the vacuum with no flux at spatial infinity. One could investigate this in zero temperature analogues of our spacetime, though such solutions have yet to be found. Thermal states, dual to bulk black holes, may instead involve a coarse-grained version of Eq.~(\ref{QNEC}) \cite{Engelhardt:2018kcs, Engelhardt:2018kzk}.

A possibility for the violation is that the bulk geometry that we have constructed is not the dominant contribution to the gravitational path integral with the prescribed boundary conditions. While we have not performed an exhaustive search, we believe our bulk solution is the dual to a thermal state on wormhole boundary. In particular, we expect that a strongly-coupled infinite rank gauge theory on a noncompact manifold confines only at zero temperature \cite{Aharony:2003sx}. In bulk language, an example of this is the well-known statement that planar AdS black holes always dominate in the canonical ensemble at any temperature \cite{Hawking:1982dh}. Since our spacetime asymptotically contains a black hole whose event horizon is planar, we expect it dominates over any bulk that lacks a black hole or lacks one whose event horizon is asymptotically planar.

It must be emphasized that while entanglement entropies for thermal states on noncompact manifolds are IR divergent, {\it a priori} Eq.~(\ref{QNEC}) does not require an IR regulator. In fact, we specifically chose an IR regulator in Eq.~(\ref{S:reg}) that respects the asympotically flat nature of our spacetime. This was in part motivated by the compactified black hole spacetime which featured a similar divergence and in which our choice of IR regulator respects the translation invariance of the background. On the other hand, if one considers the QNEC to be fundamental, then our work emphasizes that a modification of Eq.~(\ref{QNEC}) to account for IR effects is necessary.
\\
\\
{\it Note added.}---After this paper appeared on arXiv, a new paper \cite{Leichenauer:2018vot} dealing with the IR regulators for the QNEC appeared. The paper proposed two methods which may preserve the QNEC: One method considers the strip-shaped minimal surface instead of the half-space and the other method puts the system in a finite box. Some evidence in favor of the strip regulator is the saturation of the QNEC in an out-of-equilibrium holographic theory \cite{Ecker:2017jdw}. Furthermore, these regulators lead to a nice result that the variation in entanglement entropy is proportional to the thermal entropy density, $\delta \mathcal{A}/\delta r_{b} =-2\pi s$ [Eq. (9) of \cite{Leichenauer:2018vot}].

\begin{figure}[t!]
\begin{center}
\includegraphics[width=.3\textwidth]{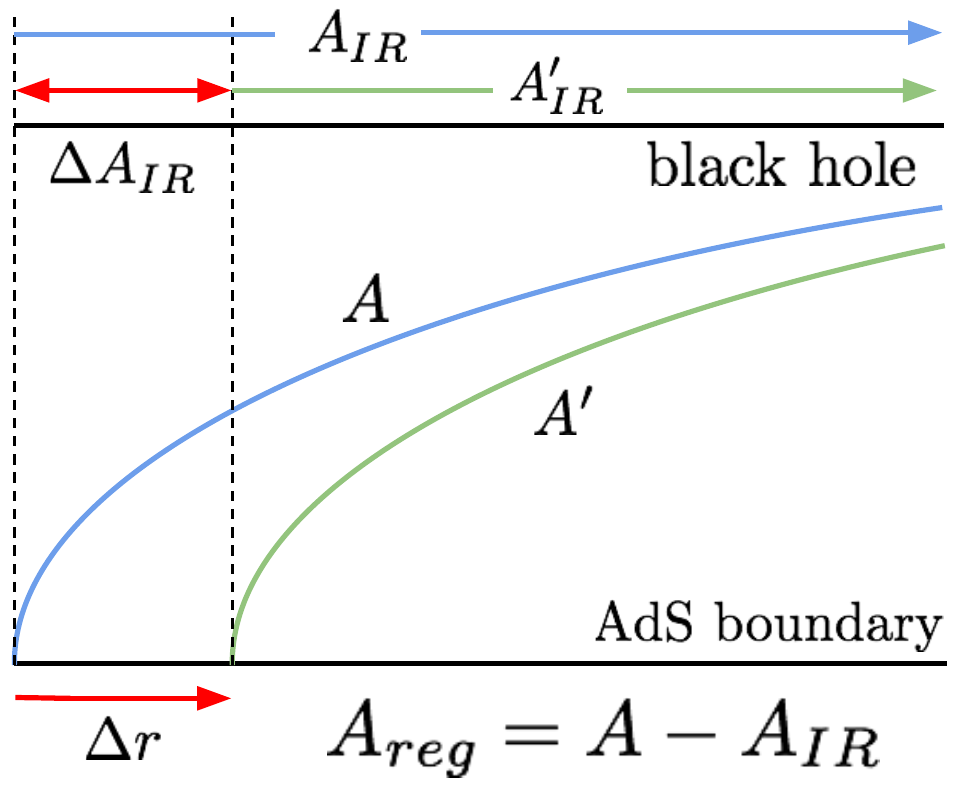}
\caption{\label{regulatorIR} Recovering $\Delta A_{reg} = -2\pi s \Delta r$ as in \cite{Leichenauer:2018vot}.}
\end{center}
\end{figure}

While these methods ensure Eq.~(\ref{UVvariation}) for the compactified black brane, it is not obvious that they accurately capture the half-space entanglement entropy since they lead to different minimal surfaces. Furthermore, for the compactified black brane, such regulators violate translation invariance on the boundary. On the other hand, in such a translationally invariant spacetime, we can derive the same result of Eq. (9) of \cite{Leichenauer:2018vot} using our regulator prescription, also illustrated in Figure~\ref{regulatorIR},

\begin{align}
A'-A_{IR}-(A-A_{IR})&=A'-(A'_{IR}+\Delta A_{IR})-(A-A_{IR})\nonumber\\
&=
(A-A_{IR})'-(A-A_{IR})-\Delta A_{IR}\nonumber\\
&=-\Delta A_{IR}=-2\pi s\Delta r
\end{align}

The two minimal surfaces are related by a diffeomorphism, $r\to r+\Delta r$, but such a diffeomorphism also moves the starting point for our regulator, $A_{IR}\to A_{IR}'$. Hence, the change in regulated area is equal to the change in the regulator area and we recover Eq. (9) of \cite{Leichenauer:2018vot}.

It is not entirely clear if the QNEC in Eq.~(\ref{QNEC}) could be preserved by the methods of Ref.~\cite{Leichenauer:2018vot} when the spacetimes in question admit no translation invariance (as in the present paper). Nevertheless, our work points to the same conclusion---the QNEC must be modified to include IR effects.  In contrast to \cite{Leichenauer:2018vot}, the IR-regulator in our proposal does not affect the functional derivative, and the regularized entropy in Eq.~(\ref{S:reg}) provides a natural physical interpretation in that it isolates the purely entanglement part of the entropy.
\\
\section{Acknowledgements}
We would like to thank G. Horowitz, D. Marolf, A. Shahbazi-Moghaddam, T. Takayanagi, and W. van der Schee for useful discussions. 
This work was supported in part by JSPS KAKENHI Grant Number 17K05451~(KM), 15K05092~(AI),  NSF Grant Number PHY-1504541 and the European Research Council 
(ERC) under the European Union’s Horizon 2020 research and innovation 
programme (grant agreement No 758759)~(EM).


\appendix



\begin{thebibliography}{99}
\bibitem{singularity_theorem}
S.~W.~Hawking and G.~F.~R.~Ellis, {\it The Large Scale Structure of Space-time} 
(Cambridge University Press, Cambridge, 1973).  

\bibitem{Hawking1971}
R.~Penrose, 
Phys.~Rev.~Lett. {\bf 14}~(1965) 57. 

\bibitem{EGJ1965} 
H.~Epstein, V.~Glaser, and A.~Jaffe, 
Nuovo Cimento {\bf 36}~(1965) 1016. 

\bibitem{GrahamOlum2007}
N.~Graham and K.~D.~Olum, 
Phys.~Rev.~D. {\bf 76}~(2007) 064001.

\bibitem{FewsterGalloway2011}
C.~Fewster and G.~Galloway, 
Classical Quantum Gravity {\bf 28}, 125009 (2011). 

\bibitem{FSW1993}  J.~L.~Friedman, K.~Schleich and D.~M.~Witt, 
Phys.~Rev.~Lett. {\bf 71}~(1993) 1486.

   \bibitem{Bousso:2015mna} 
  R.~Bousso, Z.~Fisher, S.~Leichenauer and A.~C.~Wall,
  Phys.\ Rev.\ D {\bf 93}, no. 6, 064044 (2016)
 
   \bibitem{Bousso:2015wca} 
  R.~Bousso, Z.~Fisher, J.~Koeller, S.~Leichenauer and A.~C.~Wall,
  Phys.\ Rev.\ D {\bf 93}, no. 2, 024017 (2016)
 
  \bibitem{Koeller:2015qmn} 
  J.~Koeller and S.~Leichenauer,
  Phys.\ Rev.\ D {\bf 94}, no. 2, 024026 (2016)
  
  
      \bibitem{Fu:2017evt} 
  Z.~Fu, J.~Koeller and D.~Marolf,
  Classical Quantum Gravity  {\bf 34}, no. 22, 225012 (2017)
  Erratum: [Classical Quantum Gravity  {\bf 35}, no. 4, 049501 (2018)]

    \bibitem{Maldacena:1997re}
  J.~M.~Maldacena,
  Adv.\ Theor.\ Math.\ Phys.\  {\bf 2}, 231 (1998)
  [Int.\ J.\ Theor.\ Phys.\  {\bf 38}, 1113 (1999)]

\bibitem{GJW2017}
P.~Gao, D.~L.~Jafferis, and A.~C.~Wall,
J. High Energy Phys. {\bf 12}, (2017) 151. 

\bibitem{MaldacenaMaoz04}
 J.~M.~Maldacena and L.~Maoz, 
 J. High Energy Phys. {\bf 02}, (2004) 053. 

  \bibitem{Ryu:2006bv} 
  S.~Ryu and T.~Takayanagi,
  Phys.\ Rev.\ Lett.\  {\bf 96}, 181602 (2006)
  
  \bibitem{Hubeny:2007xt} 
  V.~E.~Hubeny, M.~Rangamani and T.~Takayanagi,
  J. High Energy Phys. {\bf 0707}, 062 (2007)

%
\bibitem{Fu:2017lps} 
  Z.~Fu, J.~Koeller and D.~Marolf,
  Class.\ Quant.\ Grav.\  {\bf 34}, no. 17, 175006 (2017)
  
  \bibitem{Emparan:1999gf} 
  R.~Emparan,
  J. High Energy Phys. {\bf 9906}, 036 (1999)
  
\bibitem{Birmingham:1998nr} 
  D.~Birmingham,
 Classical Quantum Gravity  {\bf 16}, 1197 (1999)

    \bibitem{Headrick:2009pv} 
  M.~Headrick, S.~Kitchen and T.~Wiseman,
  Classical Quantum Gravity {\bf 27}, 035002 (2010)
  
  \bibitem{Figueras:2011va} 
  P.~Figueras, J.~Lucietti and T.~Wiseman,
  Classical Quantum Gravity  {\bf 28}, 215018 (2011)
  
\bibitem{Santos:2012he} 
  J.~E.~Santos and B.~Way,
  J. High Energy Phys. {\bf 1212}, 060 (2012)

\bibitem{ILL}  
The expressions 
are not illuminating and are omitted.   
    \bibitem{Fischetti:2012rd} 
  D.~Marolf, W.~Kelly and S.~Fischetti,
  arXiv:1211.6347 [gr-qc].

\bibitem{deHaro:2000vlm} 
  S.~de Haro, S.~N.~Solodukhin and K.~Skenderis,
  Commun.\ Math.\ Phys.\  {\bf 217}, 595 (2001)
    
  \bibitem{Mefford:2016res} 
  E.~Mefford,
  J. High Energy Phys. {\bf 1709}, 006 (2017)
 
 \bibitem{Headrick:2007km} 
  M.~Headrick and T.~Takayanagi,
  Phys.\ Rev.\ D {\bf 76}, 106013 (2007)
  
    \bibitem{derivation}
  At the wormhole throat, the translation symmetry along $\partial_\phi$ says $2\pi\int_{\p A} \sqrt{\gamma} \langle T_{kk}\rangle  = (2\pi)^2 \sqrt{1+\zeta} \langle T_{kk} \rangle|_{r=0}$, and for affine parameter, $d\lambda = dR\approx 2dr$. In other words $\frac{D^2S_{reg}}{D\lambda^2} \equiv k^a\delta_a(k^b\delta_b S_{reg})|_{r=0} = \frac{1}{4}\delta_r^2 S_{reg}|_{r=0}$. These combine to give Eq.~(\ref{QNEC2}). Notably, Q involves only UV and IR regular quantities and agrees with Eq.~(\ref{QNEC}) only at the throat where $\theta=0$.
  
  
  \bibitem{Balakrishnan:2017bjg} 
  S.~Balakrishnan, T.~Faulkner, Z.~U.~Khandker and H.~Wang,
  arXiv:1706.09432 [hep-th].
    
  \bibitem{Wall:2012uf} 
  A.~C.~Wall,
 Classical Quantum Gravity {\bf 31}, no. 22, 225007 (2014)

  \bibitem{extrinsicdefinition}
  From Eq.~(A.4) of Ref.~\cite{Fu:2017evt}, this is $K^{c}_{\;ab} \equiv - \gamma_{a}^{\;d}\gamma_{b}^{\;e}\nabla_d\gamma_{e}^{\;c}$.
  
  \bibitem{EWNexpression} From Eq.~(3.13) in Ref.~\cite{Fu:2017evt},the $\mathcal{O}(z^3)$ terms in the EWN expression are $h_{ab}^{(3)}k^ak^b + 2h_{ab}^{(0)}k^a\partial_\lambda\left(X^{(3)}\right)^b$ where $(X^{(3)})^b$ is the $\mathcal{O}(z^3)$ of the expansions $X^t(z) = t_{b}$ and $X^r(z) = r_{b}+\frac{K}{2} z^2 + \frac{c_1}{3}z^3 + \mathcal{O}(z^4)$. Combined with Eq.~(\ref{regularizedstresstensor}), we get $ \frac{8}{3}\left[2\pi G_4 \langle T_{kk} \rangle + \frac{h_{rr}^{(0)}}{4}k^r \partial_\lambda c_1\right] \equiv \frac{8}{3}E$.
  
  \bibitem{Leichenauer:2018obf} 
  S.~Leichenauer, A.~Levine and A.~Shahbazi-Moghaddam,
  arXiv:1802.02584 [hep-th].
  
\bibitem{Engelhardt:2018kcs} 
  N.~Engelhardt and A.~C.~Wall,
  arXiv:1806.01281 [hep-th].
  
\bibitem{Engelhardt:2018kzk} 
  N.~Engelhardt and S.~Fischetti,
  arXiv:1805.08891 [hep-th].
  
  \bibitem{Aharony:2003sx} 
  O.~Aharony, J.~Marsano, S.~Minwalla, K.~Papadodimas and M.~Van Raamsdonk,
  Adv.\ Theor.\ Math.\ Phys.\  {\bf 8}, 603 (2004)
  
  \bibitem{Hawking:1982dh} 
  S.~W.~Hawking and D.~N.~Page,
  Commun.\ Math.\ Phys.\  {\bf 87}, 577 (1983).
  
\bibitem{Leichenauer:2018vot} 
  S.~Leichenauer,
  arXiv:1808.05961 [hep-th].
  
\bibitem{Ecker:2017jdw} 
  C.~Ecker, D.~Grumiller, W.~van der Schee and P.~Stanzer,
  Phys.\ Rev.\ D {\bf 97}, no. 12, 126016 (2018)



\end{thebibliography}
\end{document}